\documentclass[aps,preprint,superscriptaddress]{revtex4-1}
\usepackage{amsmath}
\usepackage{bm}
\usepackage{graphicx}
\usepackage{xcolor}

\begin{document}

\title{Magnetic-Field-Induced Wigner Crystallization of Charged Interlayer Excitons in van der Waals Heterostructures}

\author{Igor V.~Bondarev}
\email[Corresponding author email:~]{ibondarev@nccu.edu}
\affiliation{Department of Mathematics \& Physics, North Carolina Central University, Durham, NC 27707, USA\\~\\}
\author{Yurii E.~Lozovik}
\affiliation{Institute of Spectroscopy, Russian Academy of Sciences, 142190 Troitsk, Moscow Region, Russia}
\affiliation{National Research University "Higher School of Economics",\vskip-0.05cm Tikhonov Moscow Institute of Electronics \& Mathematics, 123458 Moscow, Russia\\~\\}

\begin{abstract}
We develop the theory of the magnetic-field-induced Wigner crystallization effect for charged interlayer excitons (CIE) discovered recently in transition-metal-dichalcogenide (TMD) heterobilayers~\cite{Lius-PKim19}. We derive the ratio of the average potential interaction energy to the average kinetic energy for the many-particle CIE system subjected to the perpendicular magnetic field of an arbitrary strength, analyze the weak and strong field regimes, and discuss the 'cold' crystallization phase transition for the CIE system in the strong field regime. We also generalize the effective $\mbox{g}$-factor concept previously formulated for interlayer excitons~\cite{Nagler17}, to include the formation of CIEs in electrostatically doped TMD heterobilayers. We show that magnetic-field-induced Wigner crystallization and melting of CIEs can be observed in strong-field magneto-photoluminescence experiments with TMD heterobilayes of systematically varied electron-hole doping concentrations. Our results advance the capabilities of this new family of transdimensional quantum materials.
\end{abstract}

\maketitle

Coherent states of semiconductors with high excitation density have been at the forefront of semiconductor materials research since its inception back in the early sixties~\cite{KeldyshKozlov68,LozovikYudson,Ogawa90,LozovikPRL07,Kotthaus13,Kezer14,Fogler14,Suris16,LozovikUFN18,Lozovik19,JonFinley,FWang20,FWang21,MakShan21,CommPhys-bond}. Nowadays, a new family of ultrathin quasi-2D van der Waals (vdW) materials of controlled thickness (also called transdimensional materials~\cite{BoltSha19,BondMoSha20}), particularly bilayer transition metal dichalcogenide (TMD) heterostructures~\cite{MakShan16}, provide unique opportunities to study such states experimentally. Recent experiments have revealed that not only neutral interlayer excitons (IE) and biexcitons can form of electrons ($e$) in one layer and holes ($h$) in another layer of these bilayers due to the dimensionality reduction~\cite{Rivera2015,Ross2017,Baranowski2017,Miller17,BondVlad18}. Predicted theoretically~\cite{BondVlad18,Drummond18,Thygesen18,SnokeBond21} and then discovered experimentally~\cite{SnokeBond21,Lius-PKim19,Geim20,Xu21} are also bound three-particle fermion and four-particle boson $e$-$h$ complexes of nonzero charge such as charged interlayer excitons (CIE, or interlayer trion, a charge carrier bound to an IE~\cite{Lius-PKim19,Geim20,Xu21}) and doubly charged quaternions (two like-charge carriers bound to an IE~\cite{SnokeBond21}). Both of them feature the permanent electric dipole moment directed perpendicular to the layers and so can be effectively manipulated by an external electromagnetic field. In this way, the CIEs enable the optical control of spin-1/2 states they carry, which is advantageous for applications such as spintronics and quantum information processing, while the multiparticle collective (Bose-condensed) quaternion states offer no less than a new path to superconductivity without Cooper pairing due to their nonzero charge. For TMD bilayers under intense irradiation, therefore, collective cooperative phenomena such as crystallization and condensation phase transitions must be studied for these bound few-particle $e$-$h$ complexes theoretically and experimentally to advance potential capabilities of this new family of quantum materials.

\begin{figure}[t]
\includegraphics[scale=0.57]{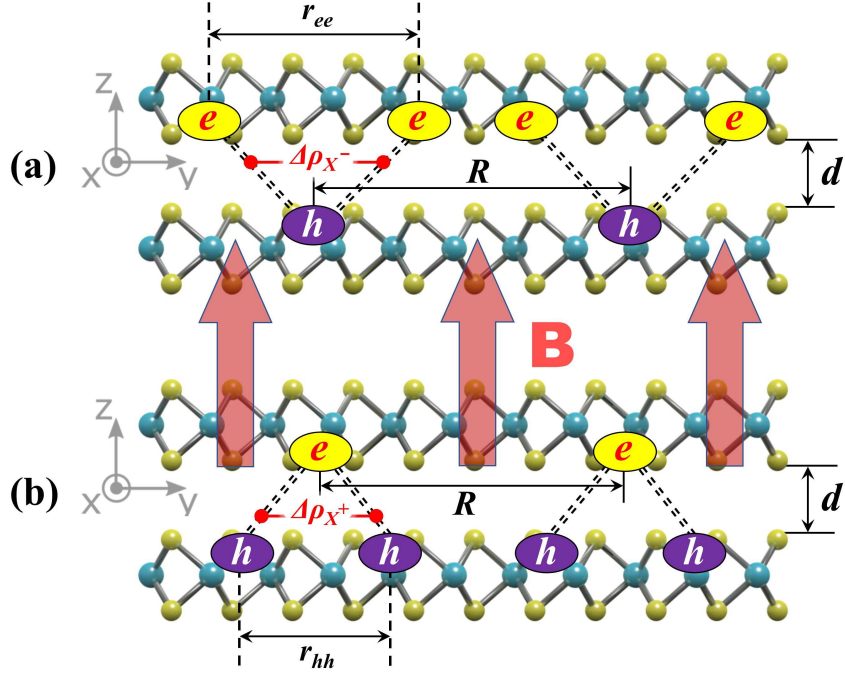}
\caption{Schematic of the bilayer TMD structure under study, with negative~(a) and positive~(b) charged interlayer exciton complexes (CIE, or interlayer $X^\pm$-trions) subjected to a perpendicular homogeneous magnetostatic field. The distance $\Delta\rho_{X^\pm}$ is between the centers of mass of the $e$-$h$ pairs forming the CIE.}
\label{fig1}
\end{figure}

Recently, there were two scenarios predicted for the long-range ordered crystal phases of the collective multiparticle CIE system in TMD bilayer nanostructures in the absence of an external magnetic field~\cite{CommPhys-bond}. They are the unlike-charge CIE crystallization and the Wigner crystallization of the like-charge CIEs, which can be selectively realized by fabricating TMD bilayers with appropriately chosen $e$-$h$ effective mass ratio and interlayer distance in addition to using the standard electrostatic doping technique. The unlike-charge CIEs form the periodic 1D chains of the two interpenetrating sublattices of positive and negative CIEs with collinear permanent dipole moments in each of the two, whereas the dipole moments of the sublattices are anticollinear. The like-charge CIEs form the 2D Wigner crystal (WC) lattice instead provided that temperature is below and the CIE density is above their respective critical values, with the latter controlled by the interlayer distance. The WC phase has been one of the longest anticipated exotic correlated phases originally thought of as a periodic array of electrons held in place when their Coulomb repulsion energy exceeds the Fermi and thermal fluctuation energies~\cite{Wigner34,Platzman74,Morf79}. Lozovik and Yudson were the first to predict the magnetic-field-induced crystallization of two-dimensional (2D) electron gas~\cite{LozYud75}, having noticed that the strong perpendicular magnetic field should turn the in-plane translational electron motion into small-radius circular orbits, which the Coulomb repulsion will then be ordering on a periodic lattice to minimize the Helmholtz free energy of the system. The Lindemann's melting criterion~\cite{Khrapak20} and the simplest single-mode (Einstein) model of lattice vibrations were used to derive the upper bound of $\sim\!100$~kG (10~T) for the critical magnetic field for the 'cold' melting ($T\!=\!0$~K) of the Wigner solid of 2D electrons~\cite{LozYud75}. The more realistic Debye vibrational model, still based on the Lindemann's melting criterion, was used to study the $T$-dependence of the stability regions of the magnetic-field-induced WC~\cite{MusLoz81}. The model was then extended to include the lattice next-nearest-neighbor anharmonicity effects based on the properly modified Lindemann's melting criterion~\cite{BedGaLoz85pla}, whereby the crystal phase shear modulus was shown to suddenly drop at melting~\cite{LozFarAbd85} and the melting phase diagram was obtained~\cite{LozFar85}. The Wigner lattice instability point was shown to be very close to the point of the Kosterlitz-Thouless topological transition in 2D systems~\cite{Kosterlitz73}, in agreement with both earlier numerical studies~\cite{Morf79} and most current thermodynamical interpretation of the Lindemann's law~\cite{Khrapak20}. The microscopic anharmonic theory of melting is confirmed by molecular dynamics simulations~\cite{BedGaLoz85jetp} and is consistent with experimental studies~\cite{GrimesAdams79,GrimesAdams80,Mendez,Kajita,Monarkha,Shayegan19,Shayegan20}. Recent experimental measurements of the magnetic-field-induced Wigner crystallization in semiconductor 2D electron systems can be found in Refs.~\cite{Shayegan19,Shayegan20}.

Here, we present our theory for the Wigner crystallization of like-charge CIEs, which is induced by the strong external magnetostatic field directed perpendicular to the TMD bilayer structure. The sketch of the system is shown in Fig.~\ref{fig1}. With magnetic field treated nonperturbatively, we derive the ratio of the potential interaction energy to the kinetic energy that controls the crystallization process for the multiparticle CIE system in the presence of the external perpendicular magnetostatic field. We analyze the weak and strong field strength limits and obtain the critical temperature for the 'cold' crystallization phase transition in the strong field regime where only the lowest Landau level is occupied. We also explain how the effects of crystallization and melting can be detected in photoluminescence (PL) experiments. We show that they can be clearly seen in magneto-PL spectra of bilayer TMD heterostructures under electrostatic doping. Finally, we remark on the universality of the Wigner crystallization phenomenon for neutral and charged quasiparticle excitations of bosonic and fermionic nature.

\subsection{Zero magnetic field like-charge trion Wigner crystallization}

In the absence of an external field, an ensemble of repulsively interacting particles (or quasi\-particles, structureless or compound) forms a WC lattice when its average potential interaction energy $\langle V\rangle$ exceeds average kinetic energy $\langle K\rangle_0$, that is~\cite{Platzman74}
\begin{equation}
\frac{\langle V\rangle}{\langle K\rangle_0}=\Gamma_0>1.
\label{Gamma0}
\end{equation}
For like-charge CIEs (or interlayer trions denoted as $X^{^{\pm}}$ in the following) shown in Fig.~\ref{fig1}, the Coulomb repulsion at large $R$ ($\gg\!r_{ee,hh}$) is strengthened at shorter $R$ by the dipole-dipole repulsion of their collinear permanent dipole moments directed perpendicular to the hetero\-structure plane and controlled by the interalayer distance $d$ of the TMD bilayer.

For trion-trion separation distances $R$ greater than the trion size ($R\!\gg\!r_{ee,hh}$ in Fig.~\ref{fig1}), the first order power series expansion of the average repulsive trion-trion interaction potential takes the form
\begin{equation}
\langle V\rangle=\frac{1}{R}\Big(1+\frac{d^2}{R^2}\Big)=\sqrt{\pi n}\,\big(1+d^2\pi n\big),
\label{Vav}
\end{equation}
where $n\!=\!1/\pi R^2$ is the trion surface density such that $d^2\pi n\!\ll1$ for our series expansion to hold true. With $d\!\sim\!3-6\,$\AA~typical of vdW systems, one then has $n\!\ll\!10^{14}\,$cm$^{-2}$ which is always the case in experiments with TMD materials where the densities are at least two orders of magnitude lower to reduce exciton nonradiative recombination~\cite{Wang17}. Also, it was shown recently~\cite{CommPhys-bond} that $r_{ee,hh}\!\lesssim\!2\;$nm and the actual inter-trion repulsion potential rapidly decreases with distance due to screening, whereby $R/r_{ee,hh}\!\gtrsim\!3$ and our CIEs are separated in space well enough to make their description in terms of free particle occupation numbers
\begin{equation}
n_\mathbf{k}=\frac{1}{e^{\,\beta(E_\mathbf{k}-\bar{\mu})}+1}
\label{occupnumber}
\end{equation}
legitimate. Here, $\beta\!=\!1/k_BT$, $E_\mathbf{k}\!=\!\hbar^2k^2/2M$, $M\!=\!M_{X^{\pm}}$ and $\bar{\mu}$ being the trion total effective mass and chemical potential, respectively. At zero $T$, this turns into a unit-step function to give $n$ in Eq.~(\ref{Vav}) in the form
\begin{equation}
n=\frac{\langle N\rangle}{S}=\frac{2}{S}\sum_\mathbf{k}n_\mathbf{k}=\frac{2}{S}\frac{S}{(2\pi)^2}\,2\pi\!\!\int_0^{k_F}\!\!\!dk\,k=\frac{k_F^2}{2\pi}
\label{n}
\end{equation}
with $S$ and $k_F$ being the surface area of the sample and the trion Fermi-momentum, respectively, whereby the translational kinetic energy per particle takes the form
\begin{equation}
\langle K\rangle_0=\frac{2}{\langle N\rangle}\sum_\mathbf{k}E_\mathbf{k}n_\mathbf{k}=\frac{\hbar^2\pi n}{2M}=\frac{\hbar^2\big(\langle k_x^2\rangle+\langle k_y^2\rangle\big)}{2M}=\frac{\langle p_x^2\rangle+\langle p_y^2\rangle}{2M}
\label{Kav0}
\end{equation}
with $\langle p_x^2\rangle\!=\!\langle p_y^2\rangle\!=\!\hbar^2\pi n/2$. This should be added by the rotational kinetic energy contribution, $K^{\texttt{(r)}}_{\!X^{^{\pm\!}}}\!=\hbar^2l(l+1)/2I_{\!X^{^{\pm}}}$, in which $l\!=\!0,1,2,...$ is the orbital quantum number and $I_{\!X^{^{\pm}}}\!=\!m_{h,e}r_{ee,hh}^2/2$ is the moment of inertia for the CIE rotation about its permanent dipole moment direction (see Fig.~\ref{fig1}). The low-$T$ statistical averaging over $l$ leads to the characteristic rotational motion "freezing" temperature $T^{\texttt{(r)}}_{X^{^{\pm\!}}}\!=\!\hbar^2\!/k_BI_{\!X^{^{\pm}}}$~(see, e.g., Ref.~\cite{Pathria}). By analogy with hydrogen molecular ion (see, e.g, Ref.~\cite{AbersQM}) this can be written in terms of the trion binding energy as $T^{\texttt{(r)}}_{\!X^{^{+\!}}}\!\approx\!\sigma|E_{\!X^{^{+\!}}}|/k_B$ and $T^{\texttt{(r)}}_{\!X^{^{\!-\!}}}\!\approx\!|E_{\!X^{^{\!-\!}}}|/k_B\sigma$ with $\sigma\!=\!m_e/m_h$, whereby the rotational degrees of freedom are to be frozen out (at least for $\sigma$ close to unity typical of TMDs~\cite{TMDmass18}) as long as the CIEs are stable against thermal fluctuations.

With no rotational term contribution and no external fields, it is straightforward to obtain a qualitative picture of the like-charge trion Wigner crystallization by performing an analysis analogous to that reported in Ref.~\cite{Platzman74} for the 2D electron gas. In view of Eqs.~(\ref{Vav}) and (\ref{Kav0}), the ratio~(\ref{Gamma0}) now takes the form
\begin{equation}
\frac{\langle V\rangle}{\langle K\rangle_0}=\frac{2M}{\hbar^2}\left(\frac{1}{\sqrt{\pi n}}+d^2\sqrt{\pi n}\right)=\Gamma_0>1,
\label{Gamma0d}
\end{equation}
whereby it can be seen that the Wigner crystallization process of CIEs can be controlled by the TMD interlayer distance $d$, an extra parameter absent from the electron WC problem. The exact expressions for the zero-$T$ critical density $n_{c}$ and the critical temperature $T^\texttt{(W)}_{c}$ of the CIE WC phase transition come out as~\cite{CommPhys-bond}
\begin{eqnarray}
n_{cX^{\pm}}\!=\!\frac{2}{\pi a_B^{\ast2}d^2}\!\left(\!\frac{g_{\pm}\Gamma_{0}}{4d}\!\right)^{\!\!2}\!\left[
1-\frac{1}{2}\!\left(\frac{4d}{g_{\pm}\Gamma_{0}}\right)^{\!\!2}\!-\sqrt{1\!-\!\left(\frac{4d}{g_{\pm}\Gamma_{0}}\right)^{\!\!2}}\,\right]\!,\nonumber\\[0.25cm]
k_{B}T^\texttt{(W)}_{cX^{\pm}}=\frac{4Ry^\ast}{g_{\pm}\Gamma_0^2}\,,\hskip3cm\label{ncTc}\\[0.25cm]
g_{\pm}(\sigma)=\left(3+\Big\{\!\begin{array}{c}1\\[-0.25cm]2\end{array}\!\Big\}\,\sigma+\Big\{\!\begin{array}{c}2\\[-0.25cm]1\end{array}\!\Big\}\,
\frac{1}{\sigma}\right)^{\!\!-1}\!\!.\hskip1.6cm\nonumber
\end{eqnarray}
Here the 3D "atomic units"\space are used~\cite{LandauQM}, with distance and energy measured in the units of exciton Bohr radius $a^\ast_B\!\!=\!0.529\,\mbox{\AA}\,\varepsilon/\mu$ and exciton Rydberg energy $Ry^\ast\!\!=\!\hbar^2\!/(2\mu\,m_0a_B^{\ast2})\!=\!e^2\!/(2\varepsilon a_B^\ast)\!=\!13.6\,\mbox{eV}\,\mu/\varepsilon^2$, respectively, $\varepsilon$ represents the \emph{effective} average dielectric constant of the bilayer heterostructure and $\mu\!=\!m_e/(\lambda\,m_0)$ stands for the exciton reduced effective mass (in the units of free electron mass $m_0$) with $\lambda\!=\!1+m_e/m_h\!=\!1+\sigma$. The zero magnetic field quantities $n_{cX^{\pm}}$ and $T^\texttt{(W)}_{cX^{\pm}}$ were analyzed in great detail in Ref.~\cite{CommPhys-bond}.

\subsection{Non-zero field like-charge trion Wigner crystallization}

In the presence of the perpendicular magnetostatic field (see Fig.~\ref{fig1}), the non-interacting-particle energy contribution to the total energy of the system of identical charged particles with non-zero spin (the many-particle CIE system in our case) can be written as~\cite{Mahan}
\begin{equation}
\frac{\hbar^2}{2M}\sum_j\Big[\Big(k_{\!jx}+\frac{qB}{2\hbar c}\,y_j\Big)^2+\Big(k_{\!jy}-\frac{qB}{2\hbar c}\,x_j\Big)^2\Big]-\sum_j{\bm\mu}_j\cdot\textbf{B}
\label{Ham}
\end{equation}
(summation over particles). Here, $\textbf{B}=\!{\bm\nabla}\!\times\textbf{A}$ is the perpendicular homogeneous magnetic field with $\textbf{A}\!=\!(-y,x,0)B/2$ representing the vector potential in the symmetric gauge. The trion effective mass $M\,(=\!M_{X^\pm})$ in the first term (Landau energy) is the sum of the $e$-$h$ band-structure masses $m_{e,h}$. The $j$-th trion magnetic moment ${\bm\mu}_j$ in the second term (Pauli interaction) is the vector sum of those of electrons and holes it is composed of, which can be written as ${\bm\mu}_j\!=\!\mbox{g}_{X^\pm}\mu_{B\,}\textbf{s}_j$ with $\mbox{g}_{X^\pm}$ and $\textbf{s}_j$ being the trion g-factor and spin-1/2 operator, respectively, and $\mu_B$ being the Bohr magneton constant. We focus on the Landau energy term first. The Pauli interaction term will be added later.

\begin{figure}[t]
\includegraphics[scale=0.57]{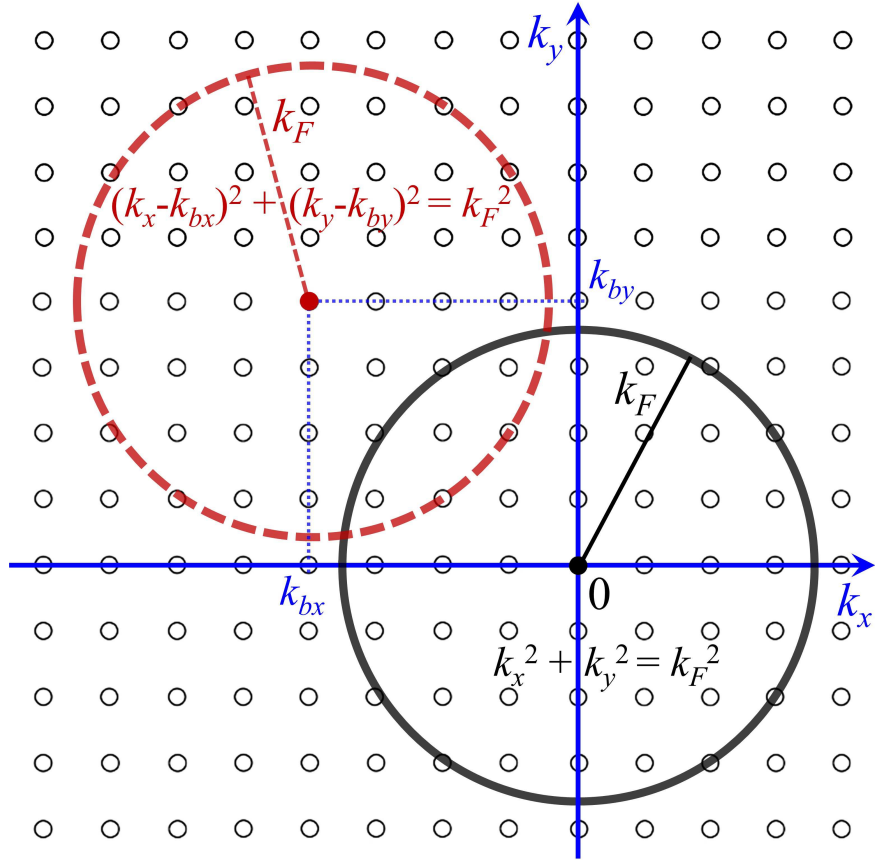}
\caption{The 2D Fermi surface center shift to the point $(k_{bx},k_{by})\!=\!(-y,x)/(2l^2)$ for the positive CIE subjected to the perpendicular homogeneous magnetostatic field $\textbf{B}$ as shown in Fig.~\ref{fig1}~(a).}
\label{fig2}
\end{figure}

Since there is no interparticle interactions in Eq.~(\ref{Ham}), the many-particle Landau energy states can be composed of the single-particle states. The single-particle Hamiltonian is
\begin{equation}
\frac{\hbar^2}{2M}\Big[\Big(k_{x}+\frac{qB}{2\hbar c}\,y\Big)^2+\Big(k_{y}-\frac{qB}{2\hbar c}\,x\Big)^2\Big].
\label{HamSP}
\end{equation}
This can be seen to include the domain of the reciprocal space $(k_x,k_y)$ inside of the 2D Fermi surface whose center is shifted out of the point $k_x\!=\!k_y\!=\!0$ due to the perpendicular magnetic field $\textbf{B}$ applied. For the many-particle Landau state at zero $T$, the average kinetic energy per particle can then be obtained by summing up over all $k$-states bounded by the displaced 2D Fermi sphere as shown for the positive CIE in Fig.~\ref{fig2} and given by
\begin{equation}
\langle K\rangle=\frac{2}{\langle N\rangle}\sum_\mathbf{k}E_\mathbf{k}n_\mathbf{k}
=\frac{2}{\langle N\rangle}\frac{S}{(2\pi)^2}\,\frac{\hbar^2}{2M}\;4\!\int_{k_{bx}}^{k_{bx}+k_F}\hspace{-0.5cm}dk_x\int_{k_{by}}^{k_{by}
+\sqrt{k_F^2-(k_x-k_{bx})^2}}\hspace{-1.75cm}dk_y\;\big(k_x^2+k_y^2\big).
\label{Kav}
\end{equation}
Here, $\langle N\rangle$ is still given by Eq.~(\ref{n}), being unaffected by the Fermi surface shift, $k_{bx}\!=\!-\pi By/\phi_0$ and $k_{by}\!=\!\pi Bx/\phi_0$ are the coordinates of the center of the 2D Fermi surface shifted, with $\phi_0\!=\!2\pi \hbar c/|q|$ representing the magnetic flux quantum. To include both positive and negative CIE case, this can generally be written as
\begin{eqnarray}
k_{bx}=-\mbox{sign}(q)\frac{\pi By}{\phi_0}=-\mbox{sign}(q)\frac{y}{2l^2}=-\mbox{sign}(q)\frac{M\omega_cy}{2\hbar}\,,\label{kbxy}\\
k_{by}=\mbox{sign}(q)\frac{\pi Bx}{\phi_0}=\mbox{sign}(q)\frac{x}{2l^2}=\mbox{sign}(q)\frac{M\omega_cx}{2\hbar}\,,\hskip0.75cm\nonumber
\end{eqnarray}
where
\begin{equation}
l^2=\frac{\hbar c}{|q|B}=\frac{\hbar}{M\omega_c}=\frac{\phi_0}{2\pi B}~~~\mbox{and}~~~\omega_c=\frac{|q|B}{Mc}=\frac{\hbar}{Ml^2}=\frac{2\pi\hbar B}{M\phi_0}
\label{lomegac}
\end{equation}
are the magnetic localization length squared and the cyclotron frequency, respectively.

The integration in Eq.~(\ref{Kav}) can be done analytically, whereupon using Eq.~(\ref{n}) to eliminate $k_F$ one obtains the exact expression as follows
\begin{equation}
\langle K\rangle=\langle K\rangle_0f(B,x,y).
\label{KavB}
\end{equation}
Here, $\langle K\rangle_0\!=\!\hbar^2\pi n/2M\!=\!\hbar^2k_F^2/4M\!=\!E_F/2$ with $E_F$ being the Fermi energy. This is the average kinetic energy per particle of Eq.~(\ref{Kav0}) with no magnetic field, the one that together with Eqs.~(\ref{Gamma0}) and (\ref{Vav}) yields the set of Eqs.~(\ref{ncTc}) reported previously~\cite{CommPhys-bond}. With Eq.~(\ref{lomegac}) this can now also be written as $\langle K\rangle_0\!=\!\hbar\omega_{c\,}l^2\pi n/2$. The dimensionless function
\begin{equation}
f(B,x,y)=1-\frac{64}{9\pi^2}+\frac{\hbar\omega_c}{2E_F}\Big[\Big(\frac{x-x_0}{\sqrt{2}l}\Big)^2+\Big(\frac{y-y_0}{\sqrt{2}l}\Big)^2\Big]
\label{fBy1}
\end{equation}
with
\begin{equation}
x_0(n,l)=-y_0(n,l)=-\mbox{sign}(q)\frac{8}{3}\sqrt{\frac{2n}{\pi}}\,l^2
\label{x0y0}
\end{equation}
is responsible for the magnetic field effect of an arbitrary strength. From Eq.~(\ref{fBy1}) the motion of particles in our system can be seen to be divided in translational and rotational parts. The latter is given by the last term describing the in-plane rotation at frequency~$\omega_c$ about the point $(x_0,y_0)$ in circles with the localization radius $\sqrt{2}\,l\,$. The magnitude of this term is controlled by its prefactor, the rotational-to-translational energy ratio. If this ratio is much less than unity, then the translational motion dominates, otherwise the rotational motion does, which corresponds to weak and strong magnetic fields, respectively. More insight into this can be gained by plugging Eq.~(\ref{x0y0}) in Eq.~(\ref{fBy1}) and regrouping the terms therein to obtain an equivalent form as follows
\begin{equation}
f(B,x,y)=1+\frac{\hbar\omega_c}{2E_F}\Big[\frac{\langle L_z\rangle}{\hbar}+\Big(\!\frac{x}{\sqrt{2}l}\Big)^{\!2}+\Big(\!\frac{y}{\sqrt{2}l}\Big)^{\!2}\Big],
\label{fBLz}
\end{equation}
whereby the rotational motion in our system can be seen more explicitly. Here,
\begin{equation}
\langle L_z\rangle=\mbox{sign}(q)\frac{16}{3\pi}\big(x\bar{p}_y-y\bar{p}_x\big)
\label{Lz}
\end{equation}
is the averaged out-of-plain angular momentum with $\bar{p}_{x,y}\!=\!\hbar\bar{k}_{x,y}\!=\!\sqrt{\langle p_{x,y}^2\rangle}\!=\!\hbar\sqrt{\pi n/2}$ being the in-plain translational momentum components statistically averaged. For weak magnetic fields $\hbar\omega_c/2E_F\!\ll\!1$ with $l\!\rightarrow\!+\infty$, which gives $f(B,x,y)\!=\!1$ in Eq.~(\ref{fBLz}) regardless of what $x$ and $y$ are. In Eq.~(\ref{fBy1}), additionally, the absolute values of $x_0$ and $y_0$ tend to infinity per Eq.~(\ref{x0y0}), yielding $(x-x_0)^2\!\sim\!x_0^2$ and $(y-y_0)^2\!\sim\!y_0^2$ with the same end result. For very strong magnetic fields, on the other hand, one has $x-x_0\!=\!l\sqrt{2}\cos\omega_ct$ and $y-y_0\!=\!l\sqrt{2}\sin\omega_ct$ with $l\!\rightarrow\!+0$, whereby both equations take the universal coordinate independent form
\begin{equation}
f(B)=1-\frac{64}{9\pi^2}+\frac{\hbar\omega_c}{2E_F}\approx\frac{\hbar\omega_c}{2E_F}=\frac{M\omega_c^2l^2}{2E_F}=\frac{1}{2l^2\pi n}=\frac{1}{\nu(n,B)}\gg1\,.
\label{fB}
\end{equation}
Here, $\nu$ is the filling factor function that determines the number of occupied states within the circular area of radius $\sqrt{2}l$ by charged particles of density $n$ bound inside~\cite{Mahan},
\begin{equation}
\nu=\pi\big(\sqrt{2}l\big)^2n=\frac{\phi_0}{B}\,n=\!\sum_{N,m_s}\!n_F(E_{N,m_s}),
\label{nu}
\end{equation}
where $n_F$ is the standard Fermi-Dirac distribution of Eq.~(\ref{occupnumber}) with
\begin{equation}
E_{N,m_s}=\hbar\omega_c\Big(N+\frac{1}{2}\Big)-\mbox{g}_{X^\pm}m_s\mu_{B}B
\label{ENms}
\end{equation}
representing the quantized energies (Landau levels) that are characterized by nonnegative integers $N\,(=\!0,1,2,...)$ corresponding to the classical rotation in quantized circular orbits. The Pauli interaction term, where $m_s\,(=\!\pm1/2)$ is the spin-1/2 projection quantum number, is hereby added for completeness. The lowest Landau level has $N\!=\!0$. All Landau states are highly degenerate as the centers of orbits can be located at any point inside the finite-size area $S\!=\!L_xL_y$ occupied by the sample. The last equality in Eq.~(\ref{nu}) can be obtained by writing the particle density $n$ in terms of the density of states $\rho(E)$ per unit area, whereby
\[
n=\!\int\!\!dE\,n_F(E)\rho(E)=\!\int\!\!dE\,n_F(E)\frac{1}{S}\sum_{\bar{k}_x=0}^{L_y/l^2}\sum_{N,m_s}\!\delta(E\!-\!E_{N,m_s})=\frac{1}{2\pi l^2}\sum_{N,m_s}\!n_F(E_{N,m_s}).
\]
This takes into account that in a finite-size sample, for the quantized $\bar{k}_x\!=\!2\pi\{0,1,2,\dots\}/L_x$ the inequality $0\!\le\!\bar{k}_xl^2\!=\!3\pi|y_0|/16\!\le\!L_y$ is to be fulfilled as per Eq.~(\ref{x0y0}) (assuming $l\!\ll\!L_y$). Plugging this in the first part of Eq.~(\ref{nu}) yields the last part.

From Eqs.~(\ref{fBLz}), (\ref{fB}) and (\ref{ENms}) it follows that the lowest Landau level $\hbar\omega_c/2$ is the only one to contribute to the function $f(B)$ in the strong field regime. In this regime, a charge carrier is localized within the area $\pi(\sqrt{2}l)^2$ due to its circular motion with out-of-plane angular momentum of absolute value $M(l\sqrt{2})^2\omega_c/2\!=\!\hbar$, as per Eq.~(\ref{lomegac}). If the magnetic localization length $l$ is much less than the interparticle distance $a$, then the repulsive interparticle interaction can promote the long-range (Wigner) crystal phase development. For the triangular lattice typical of highly correlated many-particle systems~\cite{Kleiner64}, one has $n\!=\!1/S_0$ with $S_0\!=\!a^2\sqrt{3}/2$ being the area per particle, whereby the condition $2\sqrt{2}\,l\!<\!a$ leads to an approximate upper bound $\pi\sqrt{3}/6\!\approx\!0.907$ for the filling factor $\nu$ not to exceed in order for the system to stay in the crystal phase. Rigorous microscopic quantum theories of melting yield the 2D magnetic-field-induced crystallization requirement in the form~\cite{LozFarAbd85}
\begin{equation}
\nu=\frac{\phi_0}{B}\,n=\frac{n}{n_0}=\frac{B_0}{B}\le\nu_c\approx0.13,~~~0\le\nu\le1,
\label{constraint1}
\end{equation}
mostly because of the lattice anharmonicity, for both positive and negative charge carriers regardless of their repulsive interparticle interaction strength or the crystal lattice type~\cite{endnote1}. Here, $n_0(B)\!=\!B/\phi_0\!=\!1/\pi(\sqrt{2}l)^2$ is the number of magnetic flux quanta per unit area (or the lowest Landau level
degeneracy per unit area to set up its maximum occupancy) and $B_0(n)\!=\!\phi_0 n$ is the magnetic field lower bound to be exceeded for the system to enter the strong field regime. Thus, no large density is required, the strong field is instead, for the magnetic-field-induced WC phase to occur. This conclusion can also be reached from the zero-field expressions in Eq.~(\ref{ncTc}). Plugging Eq.~(\ref{KavB}) in Eq.~(\ref{Gamma0}) redefines the energy ratio of interest to the form $\Gamma_0/f(B)\!=\!\nu\Gamma_0\!=\!\Gamma$, as per Eq.~(\ref{fB}). Solving this for $\Gamma_0$ and plugging the result in the critical density expression of Eq.~(\ref{ncTc}) yields
\[
n_{cX^{\pm}}(\nu)\approx\frac{\nu^2}{4\pi(g_\pm\Gamma)^2a_B^{\ast2}}=\frac{1}{4\pi(g_\pm\Gamma_0)^2a_B^{\ast2}}\Big(\frac{\nu}{\nu_c}\Big)^2
\]
to the lowest nonvanishing order in $\nu d/\Gamma\,(=\!\nu d/\nu_c\Gamma_0\!\ll\!1$; the $d$ dependence comes out of the next order term), whereby per Eq.~(\ref{constraint1}) no nonzero lower limit exists for $n_{cX^{\pm}}$ in the strong field regime. Similarly, for the critical temperature of Eq.~(\ref{ncTc}) one obtains
\begin{equation}
k_{B}T^\texttt{(W)}_{cX^{\pm}}(\nu)=\frac{\nu_c^2Ry^{\ast\!}}{4g_{\pm}\Gamma^2}=\frac{Ry^\ast}{4g_{\pm}\Gamma_0^2}\Big(\frac{\nu_c}{\nu}\Big)^2
=\frac{0.13^2Ry^\ast}{4g_{\pm}\Gamma_0^2}\left(\frac{B}{B_0}\right)^2,
\label{TcB}
\end{equation}
presenting $T^\texttt{(W)}_{cX^{\pm}}$ as an inverse quadratic function of $\nu$ and as a quadratic function of $B$ in the strong field regime.

\subsection{Optical detection of the magnetic-field-induced trion WC phase}

Recently, a significant enhancement of the electronic transition coupling strength to an external magnetostatic field was demonstrated experimentally for TMD heterobilayers~\cite{Nagler17}. Here, the type II (staggered gap) band alignment at stacking angles $0^\circ$ (AA stacking) or $60^\circ$ (AB stacking) enables the formation of IEs with electrons and holes in different layers. As opposed to monolayers, in TMD bilayers the valley-conserving requirement is lifted for optical $e$-$h$ transitions, allowing the IE-forming charge carriers at the $K$-points of valleys with different index (of different layers) to recombine radiatively. In an external magnetic field, their recombination yields two-component circularly polarized PL emission, which in the case of AB-stacked heterobilayers comes from the valley splitting significantly enhanced as compared to monolayers, due to the addition of conduction-band and valence-band valley magnetic moments. The effect allows for efficient external manipulation by the spin-valley degrees of freedom in the TMD heterobilayers, opening the door to new fundamental physics only achievable with these transdimensional quantum materials.

In magneto-PL experiments~\cite{Nagler17}, the TMD heterobilayer sample is first excited by linearly polarized light to equally populate both valleys of individual monolayers. Then, the optical detection and analysis follow of the two circularly polarized PL emission components ($\sigma^+$ and $\sigma^-$) resulted from the valley-selective splitting of the IE recombination transitions in the magnetic field. The total transition energy splitting is
\begin{equation}
\Delta E=E^{\sigma+}\!-E^{\sigma-}\!=\mbox{g}_\texttt{eff}\mu_BB,~~~\mbox{g}_\texttt{eff}=\pm\Big(4+\frac{2}{m_h}\mp\frac{2}{m_e}\Big),
\label{DeltaE}
\end{equation}
where masses are in units of the free electron mass, and the upper (lower) sign in the effective $\mbox{g}$-factor is for the AA-stacked (AB-stacked) heterobilayer~\cite{Nagler17}.

\begin{figure}[t]
\includegraphics[scale=0.75]{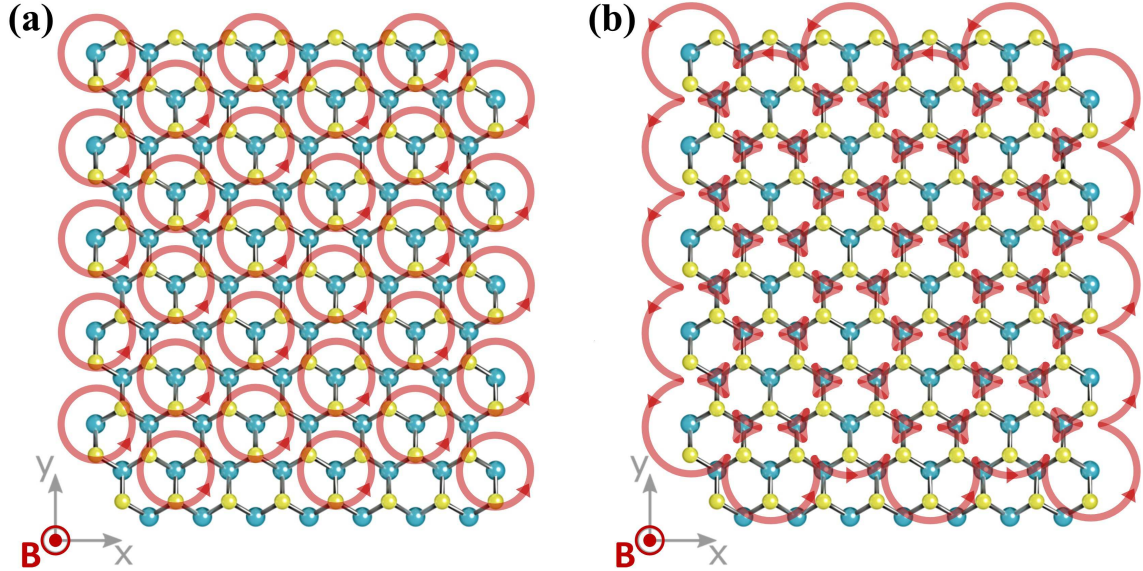}
\caption{Sketch (top view) of the crystal phase (a) and the liquid phase (b) for the negative CIEs in the strong perpendicular magnetic field (directed upward). In (a), the red periodically aligned circles with arrows show the trajectories for the negative CIE circular motion on the magnetically-induced WC lattice. In (b), the circular trajectories are increased in size to overlap, which breaks down Eq.~(\ref{constraint1}) either due to the field strength $B$ decreased or due to the doping level $n$ increased. The overlap areas are shown in white to indicate that the CIEs can now hop from circle to circle randomly, which is the liquid phase with overall rotational quasi-free CIE motion.}
\label{fig3}
\end{figure}

The TMD bilayer heterostructures can also be doped electrostatically by using two optically transparent electrical gates at the top and bottom monolayers~\cite{Lius-PKim19}. In this case, positive or negative CIEs form in the sample as per the sign of the bias voltage applied. In the sufficiently strong perpendicular magnetic field, they will be trapped by the magnetic-field-induced parabolic potential and (Wigner) crystallized due to their Coulomb repulsion. That is, the CIEs formed will be constrained to move on periodically ordered circular orbits of area $\pi(\sqrt{2}l)^2$ with angular momentum $M(l\sqrt{2})^2\omega_c/2\!=\!\hbar$, as shown in Fig.~\ref{fig3}~(a), to contribute the respective magnetic moment to $\mbox{g}_{\texttt{eff}}$ in Eq.~(\ref{DeltaE}). Increasing the doping level with magnetic field strength unchanged will initiate the melting of this WC of CIEs as prescribed by Eq.~(\ref{constraint1}). Some of the CIEs will then be forced to leave their respective Wigner lattice sites for a relatively free (quasi-free) random motion on overall circular orbits of much greater radii as shown in Fig.~\ref{fig3}~(b), being still driven by the magnetic field. Clearly, such a quasi-free motion will reduce (but will not eliminate) the $\mbox{g}_\texttt{eff}$ angular momentum contribution that comes from the CIE crystallization.

In twisted TMD bilayers, a combination of structural rippling and electronic coupling gives rise to periodic moir\'{e} potentials that can confine neutral and charged excitations and thus impede their motion~\cite{MoireIXdiff20}. At the same time, experiments have revealed that the largest moir\'{e} potentials are realized for quite a narrow range of angles that need to be controlled precisely, so that engineering of high-quality moir\'{e} lattice structures is a challenge~\cite{MoireColumbia21}. Even bilayer samples grown by chemical vapor deposition (CVD) or created by the mechanical exfoliation and transfer (MET) with a well-controlled twist angle present different results~\cite{MoireIXdiff20}. The IE diffusion length in the rotationally aligned CVD-grown heterobilayers exceeds the size of the sample (a few microns), whereas no IE diffusion is observed in the MET stacked heterobilayers with $\sim\!1^\circ$ twist angle, suggesting the IE localization by the moir\'{e} potential. Mobile excitons and trions are preferred in our case. They are more likely found in either CVD-grown samples or in MET heterobilayer samples with small moir\'{e} periods where they can tunnel between the moir\'{e} potential minima, or in commensurate homo- and heterobilayer structures with zero and near-zero moir\'{e} periods.

\begin{figure}[t]
\includegraphics[scale=0.75]{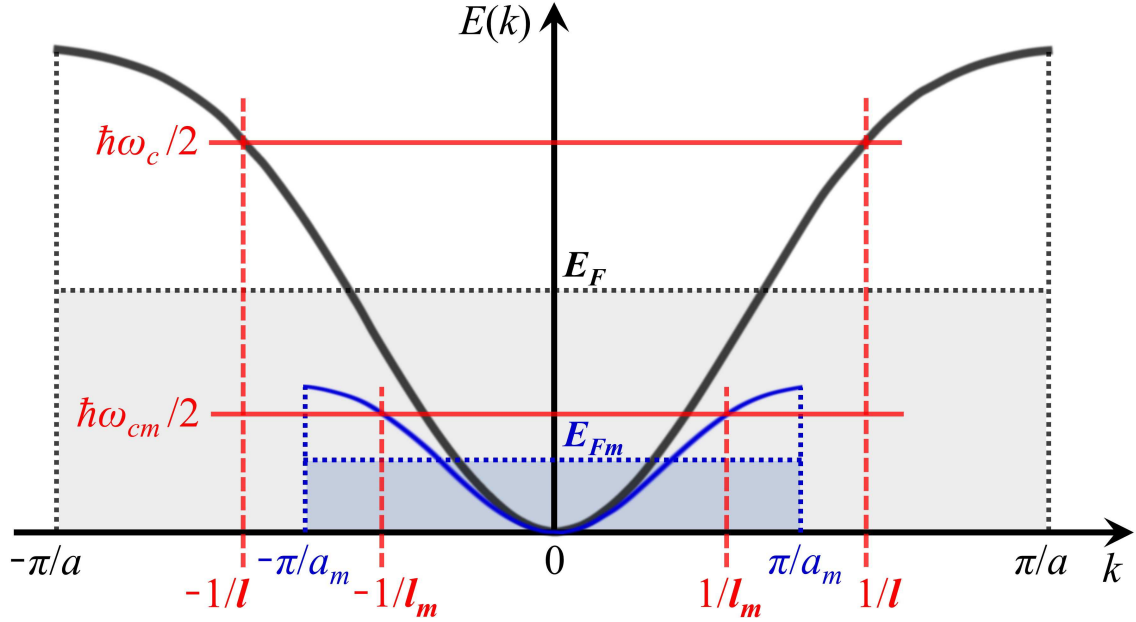}
\caption{Effective mass model sketch (not to scale) for the first Brillouin zones of the CIEs without and with periodic moir\'{e} potential present. The parabolic bands (black and blue lines, respectively) are chosen to be half-filled (shaded in grey and light blue) with $E_F$ and $E_{Fm}$ indicating their Fermi levels. The moir\'{e} potential adds to the CIE effective mass to narrow the band. The moir\'{e} pattern period $a_m$ is greater than the monolayer period $a$, to contract the zone as compared to no moir\'{e} case. The lowest Landau levels (red lines) are such that $\hbar\omega_c/2E_F\!=\hbar\omega_{cm}/2E_{Fm}$ and the magnetic length $l_m$ in the presence of moir\'{e} is greater than $l$ of no moir\'{e} case.}
\label{fig4}
\end{figure}

The role of a periodic moir\'{e} potential can be qualitatively understood in terms of the effective mass model. A~moir\'{e} potential with period $a_m$ greater than the intrinsic period $a$ of individual monolayers (always the case), adds to the CIE effective mass to shrink the CIE band and also contracts the first Brillouin zone of the reciprocal space as sketched in Fig.~\ref{fig4}. Being independent of the effective mass, the ratio $\hbar\omega_c/2E_F$ remains unchanged, so that $\hbar\omega_c/2E_F\!=\!\hbar\omega_{cm}/2E_{Fm}$, see Fig.~\ref{fig4}, but the respective magnetic length $l_m$ increases and the critical field $B_{0m}(n_m)\!=\!\phi_0n_m$ drops to keep Eq.~(\ref{fB}) constant. This leads to the $T_c$ increase as per Eq.~(\ref{TcB}). If $B\!<\!B_{0m}$ (weak field regime), then $\hbar\omega_{cm}/2E_{Fm}\!<\!1$ and the narrow-band translational motion is dominant for the CIEs trapped by the periodic moir\'{e} potential to form a 'generalized' Wigner crystal similar to that observed for electrons in twisted TMD bilayers recently~\cite{FWang20,FWang21}. If $B\!>\!B_{0m}$ (strong field regime), then $\hbar\omega_{cm}/2E_{Fm}\!>\!1$ and the rotational motion dominates for the CIEs about the centers of their respective moir\'{e} spots, provided though that the number of magnetic flux quanta $\alpha\!=\!S_{0m}B/\phi_0\!=\!S_{0m}/\pi(\sqrt{2}l_m)^2$ traversing the unit cell of the moir\'{e} lattice is not integer (the Hofstadter parameter~\cite{Hofstadter}), where $S_{0m}\!=\!a_m^2$ and $a_m^2\sqrt{3}/2$ for square and triangular moir\'{e} lattices, respectively (see also Refs.~\cite{Azbel64,ClaroWannier79,ChangNiu96} for advanced theory and Ref.~\cite{FrancoNori16} for review of recent experimental work). The magnetic-field-induced CIE Wigner crystallization is to proceed as prescribed by Eq.~(\ref{constraint1}) with $B_0$ replaced by $B_{0m}$ in this case. If $\alpha$ is integer, then the magnetic field effect vanishes identically~\cite{Hofstadter}, whereby the CIEs form a narrow-band 'generalized' Wigner crystal on the moir\'{e} lattice, with no rotation present in the system and so no $\mbox{g}_\texttt{eff}$ change in Eq.~(\ref{DeltaE}). An example we consider below is the commensurate AA-stacked MoSe$_2$/WSe$_2$ heterobilayer system studied in Ref.~\cite{Lius-PKim19}.

It has been recently shown experimentally~\cite{Xu21} that the g-factors obtained using Eq.~(\ref{DeltaE}) are indistinguishable for IEs and CIEs. Indeed, the Zeeman effect of magnetic field on the trion energy is given by the Pauli interaction term in Eq.~(\ref{ENms}), and can be viewed as the sum of the IE Zeeman shift and that of the extra carrier. Clearly, the extra electron (hole) in a negative (positive) trion only shifts the transition energy~\cite{Xu21}, but does not contribute to the spectroscopic Zeeman splitting given by Eq.~(\ref{DeltaE}). This does not include the strong field regime, however, where in addition to the Pauli interaction term the Landau energy term of Eq.~(\ref{ENms}) contributes the angular momentum of Eq.~(\ref{Lz}) to the effective g-factor of the entire charge system due to the circular motion of the CIE as a compound charged particle, which leads to the WC formation of CIEs provided that Eq.~(\ref{constraint1}) is fulfilled.

A simple phenomenological model allows one to understand the generic behavior of the effective $\mbox{g}$-factor in Eq.~(\ref{DeltaE}) in the case of the Wigner solid-liquid phase transition of the CIEs in doped TMD heterobilayers in the strong magnetic field regime. Let the CIE trapping rate $\gamma_t$ to and detrapping rate $\gamma_f$ from the magnetic-field-induced parabolic localization potential be of the standard exponential form as follows
\begin{equation}
\gamma_t=A_te^{a_tU},~~~\gamma_f=A_fe^{a_f(U-{\cal{E}})}.
\label{gamma_tf}
\end{equation}
Here, ${\cal{E}}$ stands for the CIE crystal-liquid phase internal energy difference, $U$ represents the charge carrier potential energy change with doping (proportional to the gate voltage that controls the dopant concentration), both in units of $k_BT$, and $A_{t,f}$ are the CIE trapping-detrapping amplitudes. The constants $a_{t,f}$ in the exponents are necessary to convert the charge carrier energy added into the sample with doping into that of the CIEs formed due to doping since not all of the doped charge carriers participate in the CIE formation. The statistical detailed balance further requires the condition
\[
\gamma_tB_f=\gamma_fB_te^{a_f{\cal{E}}+(a_t-a_f)U}
\]
to be fulfilled, where $B_t$ and $B_f$ are the effective numbers of trapped and quasi-free states available for quasi-free and trapped CIEs to get trapped and get quasi-free, respectively. Plugging Eq.~(\ref{gamma_tf}) in here gives $A_f\!=\!B_fA_t/B_t=CA_t$, whereby the detrapping amplitude comes out being proportional to the trapping amplitude. Next, we introduce the relative weights of the $e$-$h$ recombination components to take part in the magneto-PL emission under doping. They are, respectively, $w_t^+$, $w_f^+$, $w_h$ for trapped $X^+$, quasi-free $X^+$ and holes when doped by holes, and $w_t^-$, $w_f^-$, $w_e$ for trapped $X^-$, quasi-free $X^-$ and electrons when doped by electrons, all defined explicitly as follows
\begin{equation}
w_{t,f}^{\pm}=\frac{\gamma_{t,f}(X^\pm)}{1+\gamma_t(X^\pm)+\gamma_f(X^\pm)},~~w_{h,e}=\frac{1}{1+\gamma_t(X^\pm)+\gamma_f(X^\pm)},~~w_{h,e}+w_t^\pm+w_f^\pm=1.
\label{weights}
\end{equation}
Here, $\gamma_{t,f}(X^+)$ and $\gamma_{t,f}(X^-)$ are given by Eq.~(\ref{gamma_tf}) with $A_{t,f}$ and $a_{t,f}$ specified for the positive and negative CIEs individually. These weight functions of $\cal{E}$ and $U$ now require that $\mbox{g}_\texttt{eff}$ of Eq.~(\ref{DeltaE}) be redefined consistently to take into account the positive and negative CIEs formed under hole and electron doping, respectively. We also note that in the vicinity of the CIE crystal-liquid phase transition, taking into account Eq.~(\ref{TcB}), the following is true:
\begin{equation}
{\cal{E}}=\frac{\delta E}{k_BT}\approx\frac{\delta E}{k_BT^\texttt{(W)}_{cX^{\pm}}}=\kappa\Big(\frac{\nu}{\nu_c}\Big)^2,~~~\kappa=\delta E\,\frac{4g_{\pm}\Gamma_0^2}{Ry^\ast}\,.
\label{InternalEnergy}
\end{equation}
Here, $\delta E$ is the crystal-liquid phase internal energy difference in the physical units and $\kappa$ is the dimensionless constant. With this in view, one finally obtains the effective $\mbox{g}$-factor as a function of the two independent variables $U$ and $\nu/\nu_c$. The former is controlled by the gate voltage and varies the doped charge carrier concentration to result in the trion formation. The latter, treated as a function of $B$ only as per Eqs.~(\ref{constraint1}) and (\ref{TcB}), can be viewed as that setting up the \emph{initial} pre-doped value of the filling factor $\nu=\bar{\nu}=\!B_0/B$ for $e$-$h$ pairs created by absorption of light (and controlled by light intensity) in an \emph{undoped} TMD bilayer (with no trions formed) subjected to the strong external magnetic field.

The effective $\mbox{g}$-factor of Eq.~(\ref{DeltaE}) is now to be redefined for the hole ($\nu_h$) and electron ($\nu_e$) doping individually as follows:
\begin{equation}
\mbox{g}_\texttt{eff}^{+}=\pm\Big[4+2\Big(\frac{w_h}{m_h}+\frac{w_t^{+}+w_f^{+}}{M_{X^+}}\Big)\mp\frac{2}{m_e}\Big]
\label{geff+}
\end{equation}
with $M_{X^+}\!=\!2m_h+m_e$ for doping with holes and
\begin{equation}
\mbox{g}_\texttt{eff}^{-}=\pm\Big[4+\frac{2}{m_h}\mp2\Big(\frac{w_e}{m_e}+\frac{w_t^{-}+w_f^{-}}{M_{X^-}}\Big)\Big]
\label{geff-}
\end{equation}
with $M_{X^-}\!=\!2m_e+m_h$ for doping with electrons.

\begin{figure}[t]
\includegraphics[scale=0.85]{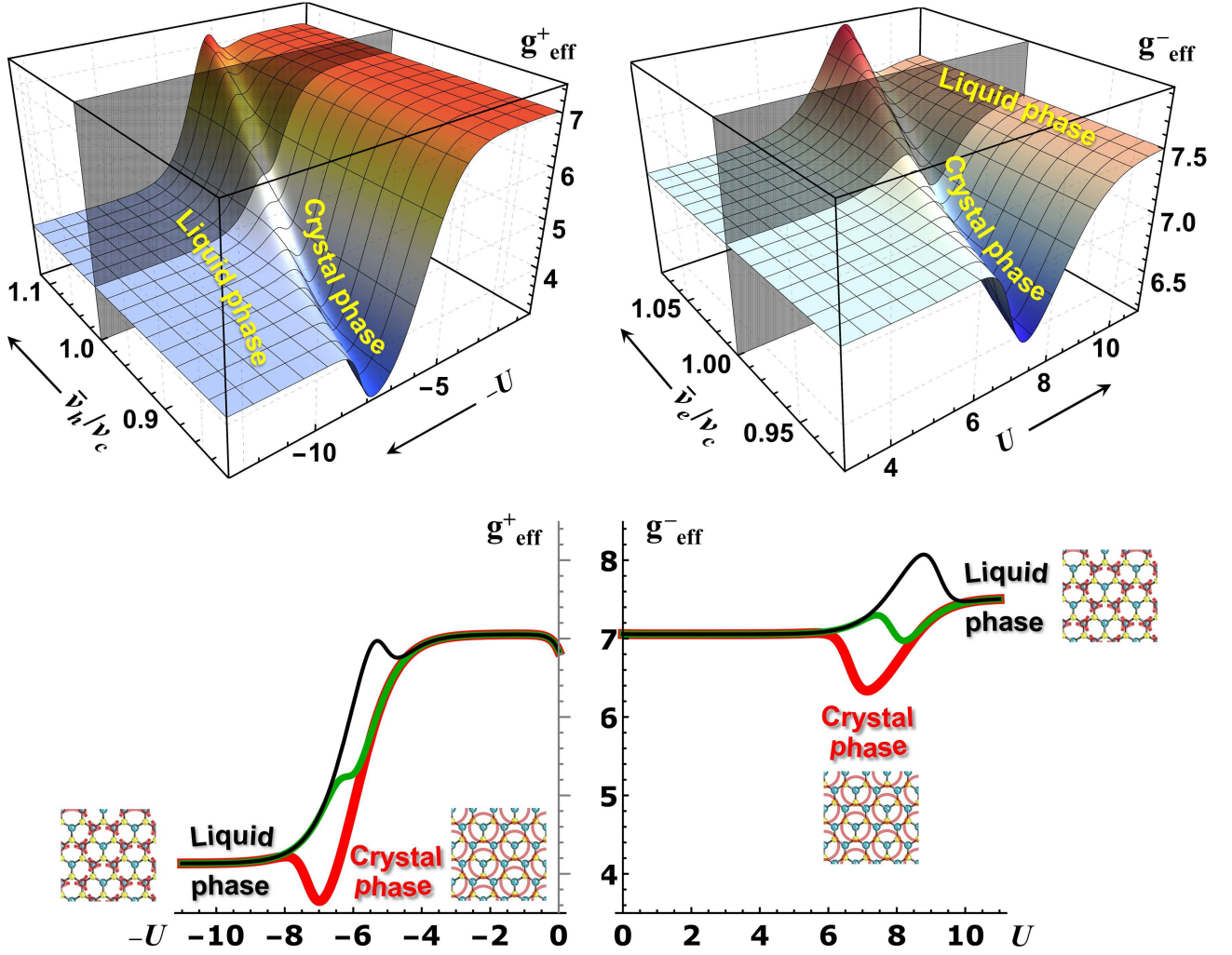}
\caption{The effective $\mbox{g}$-factor as given by Eqs.~(\ref{gamma_tf})--(\ref{geff-}) of the CIE Wigner crystallization model for the positive (top left) and negative (top right) CIEs in a generic AA-stacked TMD heterobilayer with doped carrier energy $U$ and initial filling factors $\bar{\nu}_{e,h}/\nu_c$ varied in the strong perpendicular magnetic field regime. Bottom panel shows the crosscuts of the top graphs by the vertical planes $\bar{\nu}_{e,h}/\nu_c=0.9$, $1.0$, $1.1$ (line thickness increase indicates stronger field applied). See text for details.}
\label{fig5}
\end{figure}

Figure~\ref{fig5} shows our calculations of the effective $\mbox{g}$-factor as given by the CIE Wigner solid-liquid model in Eqs.~(\ref{gamma_tf})--(\ref{geff-}) for the positive (top left panel) and negative (top right panel) CIEs in a generic AA-stacked TMD heterobilayer taken as an example. For definiteness, we use the $e$-$h$ effective masses of the MoSe$_2$/WSe$_2$ heterostructure studied experimentally in Ref.~\cite{Lius-PKim19}. They are $m_h\!=\!0.36$ for the holes of the WSe$_2$ monolayer~\cite{Kormanyos} and $m_e\!=\!0.8$ for the electrons of the MoSe$_2$ monolayer~\cite{Larentis18}. This gives $\mbox{g}_\texttt{eff}\!\approx\!7$ for the undoped heterobilayer described by Eq.~(\ref{DeltaE}), in which case no CIEs are present in the system. Other dimensionless model parameters we use are chosen to be consistent with the detailed balance condition above (to be obtained from fitting our model to the experimental data when available). They are as follows: $A_t\!=\!4\times10^{-5}$, $a_t\!=\!1.7$, $A_f\!=\!10^3A_t$, $a_f\!=\!3a_t$, $\kappa\!=\!5.5$ for the positive CIE and $A_t\!=\!10^{-6}$, $a_t\!=\!1.7$, $A_f\!=\!10^3A_t$, $a_f\!=\!3a_t$, $\kappa\!=\!6.5$ for the negative CIE. The graphs in both panels can be seen to develop the valleys as $U$ departs from zero with doping under increasing gate voltage, whereby the CIEs start forming at small concentrations and getting crystallized as long as the condition $\nu_{h,e}\!<\!\nu_c$ (or $\nu_{h,e}/\nu_c\!<\!1$) is fulfilled for increasing $\nu_{h,e}$ as they depart from their initial values $\bar{\nu}_{e,h}$ per Eq.~(\ref{constraint1}). The above-mentioned $\kappa$ parameters have been adjusted to position the planes $\bar{\nu}_{e,h}/\nu_c\!=\!1$ (light grey color in the figure) right where these 'crystallization' valleys start developing with $\bar{\nu}_{h,e}$ decreasing, which allows one to evaluate $\delta E$ of Eq.~(\ref{InternalEnergy}). With $e$-$h$ effective mass ratio $\sigma\!=\!m_e/m_h\!=\!0.8/0.36\!\approx\!2.2$, Eq.~(\ref{ncTc}) gives $g^+\!\approx\!1/6.3$ and $g^-\!\approx\!1/7.9$, whereby solving Eq.~(\ref{InternalEnergy}) for $\delta E$ one obtains $5.5\times6.3\!\approx\!35$ and $6.5\!\times7.9\!\approx\!51$ in units of $Ry^\ast/4\Gamma_0^2$ for $X^+$ and $X^-$, respectively. Next, using $Ry^\ast\!\approx\!0.1$~eV recently reported for TMDs~\cite{CommPhys-bond} and $\Gamma_0\!=\!55$ reported previously for 2D parallel dipole systems from molecular dynamics simulations~\cite{BedGaLoz85jetp}, we arrive at $\delta E\!\approx\!0.3$~meV for $X^+$ and $\delta E\!\approx\!0.4$~meV for $X^-$, which is consistent with earlier numerical studies of the Wigner crystallization phenomena in 2D $e$-$h$ systems~\cite{BedGaLoz85jetp}. To give an idea of temperatures, densities and fields involved in our case, we start with $n\!\approx\!5\!\times\!10^{11}$~cm$^{-2}$ reported for IEs in the experiments with MoSe$_2$/WSe$_2$ heterostructures~\cite{Lius-PKim19}. The CIE density is expected to be significantly lower than that since only a fraction of doped charge carriers couples to IEs to form CIEs. Taking $n\!\sim\!10^{10}$~cm$^{-2}$ for CIEs, Eq.~(\ref{constraint1}) yields $B_0\!\approx\!0.4$~T requiring that $B\!>\!B_0/\nu_c\!\approx\!0.4/0.13\!\approx\!3$~T for the WC phase of CIEs to occur. If $B\!=\!45$~T, for example, then $B/B_0\!=\!\nu_c/\nu\!=\!15$ to give $T^\texttt{(W)}_{cX^+}\!\approx\!2$~K and $T^\texttt{(W)}_{cX^-}\!\approx\!3$~K as per Eq.~(\ref{TcB}) with other parameters as mentioned above. These values are all practically achievable in real experiments.

If the initial values $\bar{\nu}_{h,e}$ are less than $\nu_{c\,}$, then with $U$ further departing from zero the CIE densities increase together with their respective doped charge carrier densities, pushing the \emph{current} $\nu_{h,e}$ values closer and closer to the $\nu_c$ value to finally exceed it, which breaks the WC phase condition of Eq.~(\ref{constraint1}). This is why in the top left and top right panels of Fig.~\ref{fig5} after passing the crystallization drop-off the effective $\mbox{g}$-factors raise up and plateau at values quite different from $\mbox{g}_\texttt{eff}\!\approx\!7$, the no-doping value. The CIE WC melts here as the crystallization condition (\ref{constraint1}) is no longer met when the \emph{then-current} $\nu_{h,e}$ values start exceeding the $\nu_c$ value. To give a better view of how this happens, at the bottom of Fig.~\ref{fig5} we show the crosscuts of the top graphs by the vertical planes $\bar{\nu}_{e,h}/\nu_c\!=\!0.9$, $1.0$, and $1.1$, where thicker lines indicate smaller $\bar{\nu}_{e,h}/\nu_c$ (or stronger magnetic fields applied). For $\bar{\nu}_{e,h}/\nu_c\!=\!0.9$ the trion liquid phase follows the WC phase with $U$ increasing under doping, whereas for $\bar{\nu}_{e,h}/\nu_c\!=\!1.0$ and $1.1$ the WC phase does not occur and trions formed under charge doping enter the liquid phase directly. Either of these scenarios will always be the case for the initially undoped bilayer system in the strong field regime with charge carrier doping level smoothly increasing from zero. They should in principle be distinguishable in the CIE transport measurements similar to those reported in Ref.~\cite{Lius-PKim19}, not only in the magneto-PL measurements we are focusing on here. As opposed to the CIE liquid phase, the WC phase is not expected to show in-plane current due to being pinned on circular orbits inside the sample [cf. Fig.~\ref{fig3}~(a) and Fig.~\ref{fig3}~(b)]. If, on the other hand, the bilayer system is first sufficiently doped to form the CIEs and then is subjected to a smoothly increasing perpendicular magnetic field, neither of the two CIE phases will occur until the field gets strong enough to approach the Wigner crystallization condition. Thus, the strong-field magneto-PL experiments with TMD heterobilayes of systematically varied $e$-$h$ doping concentrations make it possible to study this unique strongly-correlated many-particle phenomenon of CIE Wigner crystallization.

\subsection{Conclusion and final remarks}

We develop the theory for the magnetic-field-induced Wigner crystallization of charged interlayer excitons that were discovered recently in TMD heterobilayers~\cite{Lius-PKim19}. We derive the ratio of the average potential interaction energy to the average kinetic energy for the many-particle CIE system subjected to the perpendicular magnetic field of an arbitrary strength at zero temperature. We analyze the weak and strong field regimes and obtain the critical temperature for the 'cold' crystallization phase transition in the strong field regime where the lowest Landau level alone is occupied. We also generalize the effective $\mbox{g}$-factor concept previously formulated for interlayer excitons~\cite{Nagler17}, to include the formation of CIEs under electrostatic doping of heterobilayers in the strong perpendicular magnetic field. We show that the magnetic-field-induced Wigner crystallization and melting of the many-particle system of CIEs can be detected and monitored in magneto-PL experiments that measure the effective $\mbox{g}$-factor of bilayer TMD heterostructures under controlled electrostatic doping. The CIE crystal-liquid internal energy difference is estimated to be $\sim\!0.3$~meV for a typical case of the AA-stacked MoSe$_2$/WSe$_2$ heterobilayer.

On a final note, we wish to emphasize that the magnetic-field-induced 2D Wigner crystallization effect in the strong perpendicular field is a universal phenomenon. In this regime, the only internal parameter that remains there to control the properties of the system is the cyclotron frequency, which exceeds all other internal parameters including crystal lattice vibrational frequencies and interparticle coupling strength. Hence, any many-particle (or quasiparticle) system with repulsive interparticle interactions, be it fermionic or bosonic, be it charged such as electrons, holes, CIEs, quaternions, or even neutral such as IEs, will experience the Wigner crystallization phase transition (which we will give the full account of elsewhere) when the external magnetostatic field is strong enough to isolate the lowest Landau level and make the cyclotron frequency the leading parameter of the system. That is why the Wigner crystallization condition (\ref{constraint1}) we use here is universal, being independent of the crystal lattice type and the sort of (quasi-)particles crystallized. Despite this universality, however, only for CIEs the Wigner crystallization effect can be detected and monitored in the effective $\mbox{g}$-factor measurement experiments we discuss here, as the features we predict for the $\mbox{g}$-factor are all coming from the additional magnetic moment of the \emph{charged} IEs that is associated with their rotational angular momentum in the crystallized phase.

\section{Data Availability} The data that supports the findings of this study are available within the article.

\section{Code Availability} Wolfram Mathematica code used to generate the graphs in Fig.~\ref{fig5} is available from the corresponding author upon email request.

\section{Acknowledgments}I.V.B. acknowledges discussions with Luis Jauregui (UC Irwine), Andrew Joe (Harvard), and Philip Kim (Harvard) at the beginning of this work. This research is supported by the U.S. Department of Energy, Office of Science, Office of BES under award No.$\,$DE-SC0007117 (I.V.B.) and by the RFBR grants No.$\,$20-02-00410 and No.$\,$20-52-00035 (Yu.E.L.). Yu.E.L. is supported by the Basic Research Program at the National Research University HSE. I.V.B. acknowledges hospitality of the Kavli Institute for Theoretical Physics (KITP), UC Santa Barbara, where this work was completed during his invited visit as a KITP Fellow 2022--23 (the research program supported by the U.S. National Science Foundation under Grant No. PHY-1748958).

\section{Author contributions}

Both authors conceived the project. I.V.B. developed the theory, carried out numerical calculations, and wrote the final version of the manuscript. Yu.E.L. provided expertise in many-particle electron-hole correlations and crystallization phenomena. The authors discussed the results, their significance, and the ways to best represent them in the manuscript.

\section{Competing interests}

The authors declare no competing financial interests.


\begin{thebibliography}{99}
\bibitem{endnote1}We use the convention by which $\nu_e$ of electrons and $\nu_h$ of holes are counted from $0$ upward and from $1$ downward, respectively, whereby $0\le\nu_{e,h}\le1$ and $\nu_e+\nu_h=1$ as required by the electron-hole interchange symmetry.
\end{thebibliography}
\end{document}